\documentclass[reprint, superscriptaddress,amsmath,amssymb,aps,showkeys,showpacs,
twoside,final,secnumarabic,
nofootinbib]{revtex4-2}

\usepackage[paperwidth=205mm,paperheight=290mm,top=17mm,bottom=25mm,
inner=17mm,outer=17mm,
twoside]{geometry}

\usepackage{cmap} 
\usepackage[T1,T2A]{fontenc}
\usepackage[utf8]{inputenc}
\usepackage[russian,english]{babel}
\usepackage{color}
\usepackage{physics}

\usepackage{graphicx}
\usepackage{dcolumn}
\usepackage{subfigure}

\usepackage{bm} 
\usepackage[unicode=true,colorlinks=true,linkcolor=magenta, urlcolor=blue, citecolor = blue,breaklinks]{hyperref}
\usepackage{multirow}
\usepackage{url}
\DeclareGraphicsExtensions{.eps}

\newcount\issue
\newcount\Vol
\newcount\numb
\usepackage{fancyhdr} 
\def\Vol{\textbf{78}}
\def\numb{x}
\begin{document}

\title{Neutrino physics\\[20pt]
	Constraining axial non-standard interactions of neutrinos  \\with long baseline experiments} 

\def\addressa{	School of physics, Institute for Research in Fundamental Sciences (IPM)
	\\
	P.O.Box 19395-5531, Tehran, Iran}

\author{\firstname{Yasaman}~\surname{Farzan} }
\email[E-mail: ]{yasaman@theory.ipm.ac.ir}
\author{\firstname{Saeed}~\surname{Abbaslu}  }
\email[E-mail: ]{s-abbaslu@ipm.ir}
\affiliation{\addressa}

\received{xx.xx.2025}
\revised{xx.xx.2025}
\accepted{xx.xx.2025}

\begin{abstract}
Thanks to a number of neutrino oscillation and Coherent Elastic neutrino Nucleus Scattering (CE$\nu$NS) experiments, the vector Non-Standard Interactions (NSI) of neutrinos have been well studied and constrained. We show that the long-sought-after new physics may hide in the ``axial" non-standard interactions rather than in the vector NSI. We then show how by studying neutral current scattering events in the detectors of long baseline experiments, MINOS, MINOS$+$ and DUNE, the impact of the axial NSI can be discovered.
\end{abstract}

\pacs{Suggested PACS}\par
\keywords{Non-standard Interactions (NSI), axial interaction, neutrinos, DUNE, MINOS(+)    \\[5pt]}

\maketitle
\thispagestyle{fancy}


\section{\label{intro} Introduction}
Historically,  the establishment of $SU(2)\times U(1)$ gauge symmetry as the model underlying the electroweak interactions made possible by discovering the neutral current (NC) scattering of neutrinos off the matter fields. It is intriguing to speculate that the first glimpse of new physics may show up in the non-standard interactions of the neutrinos with matter field. In recent years, rich literature  has been developed on  the  NC NSI of neutrinos with quarks which can be parameterized as the following four-Fermi interaction term:
\begin{equation}
V_{NSI}=\frac{G_F}{\sqrt{2}}[\bar{\nu}_\alpha \gamma^\mu (1-\gamma^5)\nu_\beta][\bar{q}\gamma_\mu(\epsilon_{\alpha\beta}^{Vq}+\epsilon_{\alpha\beta}^{Aq}\gamma^5)q]  \ ,
\label{eff}\end{equation} in which 
$\epsilon_{\alpha\beta}^{Vq}$ and $\epsilon_{\alpha\beta}^{Aq}$ are vector and axial non-standard interaction couplings with $q\in \{u,d \}$.
Non-zero $\epsilon^{Vq}$ can show up in the neutrino propagation in matter and CE$\nu$NS  but, $\epsilon^{Aq}$ does not affect these processes. Null results for search of such effects have set strong bounds on $\epsilon^{Vq}$. In comparison, $\epsilon^{Aq}$ is much less studied and constrained. In particular, $\epsilon^{Aq}_{\tau \tau}$ can be  as large as 1. However, as shown in \cite{Abbaslu:2024hep}, rich new physics may lie behind $\epsilon^{Aq}_{\tau \tau}\sim 1$. Both axial and vector NSI can affect the interaction rate of (incoherent) neutrino scattering off nuclei. As a result, we can use the neutral current data of long baseline experiments such as MINOS(+) or DUNE to constrain $\epsilon^{Aq}$ \cite{Abbaslu:2023vqk,Abbaslu:2024jzo}.
Considering that due to the oscillation at the far detector of these experiments, the neutrino flux is mainly composed of $\nu_\tau$, the data can be efficient in probing $\epsilon^{Aq}_{\tau \tau}$ and $\epsilon^{Aq}_{e \tau}$.

 The NuTeV experiment has already constrained the $\mu \mu$ and $\mu \tau$ components of $\epsilon^{Aq}$ \cite{NuTeV:2001whx}.
The $e \mu$ component is even more severely constrained by searches of lepton flavor violating process, $\mu +{\rm Ti} \to e {\rm Ti}$ \cite{Davidson:2003ha}.
In the previous works \cite{Abbaslu:2023vqk,Abbaslu:2025fwt}, we 
have studied the possibility to improve these bounds by the present MINOS($+$)
and the upcoming DUNE data. Since the prospect is limited, we will not repeat those results here. We shall focus on the $ee$, $e\tau$ and $\tau\tau$ components on which the present bounds are weak. The strongest
bounds on these components come from the SNO data on the Deuterium dissociation by solar neutrinos on the Earth \cite{Coloma:2023ixt}.
This process is however sensitive only to the isospin violating interactions and can only constrain the difference between the couplings to the $u$ and $d$ quarks:
$$ -0.19<\epsilon_{ee}^{Au}-\epsilon_{ee}^{Ad}<0.13 \ \ {\rm and } \ \   -0.13<\epsilon_{e\tau}^{Au}-\epsilon_{e\tau}^{Ad}<0.1 \ $$
and
\begin{equation}-2.1<\epsilon_{\tau \tau}^{Au}-\epsilon_{\tau \tau}^{Ad}<-1.8 \ \  {\rm or} \ \  -0.2 <\epsilon_{\tau \tau}^{Au}-\epsilon_{\tau \tau}^{Ad}<0.15 \ .\label{nontrivial}
\end{equation}
Notice that along with the solutions around zero,  disconnected solutions are also found. As we shall see, the MINOS($+$) results completely rule out these solutions \cite{Abbaslu:2025fwt}. Since there are already strong bounds on the vector NSI from the neutrino oscillation experiments as well as from coherent elastic neutrino nucleus scattering  (CE$\nu$NS) experiments, we set their values equal to zero.

  The present  letter is organized as follows. In sect. \ref{model}, we present
a model leading to $\epsilon^{Au}_{\tau\tau}=\epsilon^{Ad}_{\tau\tau}\sim O(1)$.
In sect. \ref{NSI-impact}, we show how the axial NSI can affect the neutral current events at  the long baseline experiments. In sect. \ref{exps}, we describe the MINOS, MINOS$+$ and DUNE experiments. In sect. \ref{results}, we show the present bounds from MINOS($+$) and the forecast for DUNE. The results are summarized in sect. \ref{Con}.

\section{ \label{model} A model for axial NSI}
In this section, we present a model with large axial NSI; {\it i.e.,} with $\epsilon^{Aq}\sim 1$. The model, which was proposed in \cite{Abbaslu:2024hep}, is based on  
a new $U(1)$ gauge symmetry under which the left-handed and right-handed first generation quarks have opposite charges
$$ Q_1 \equiv  \left( \begin{matrix} u_L \cr d_L \end{matrix} \right) \to e^{i\alpha} Q_1$$ and
$$ u_R \to e^{-i\alpha} u_R \ \ {\rm and} \ \ d_R\to  e^{-i\alpha} d_R
\ .$$
Then, integrating out the gauge boson denoted by $Z'$, we obtain
$$\epsilon^{Au}_{\tau \tau}=\epsilon^{Ad}_{\tau \tau}=\frac{9 \sqrt{2} g_{Z'}^2}{m_{Z'}^2 G_F}$$
with vanishing other components of NSI.

As shown in \cite{Abbaslu:2024hep}, without violating any bound, $\epsilon^{Aq}_{\tau \tau}$ as large as $O(1)$ is possible within this model.
Anomaly cancellation requires a chiral dark matter and new heavy quarks charged under new $U(1)$ which can be discovered in near future. The model finds correlation between searches for new quarks in collider and dark matter direct detection with a nonzero $\epsilon^{Aq}_{\tau\tau}$ to be discovered by DUNE. For more details, see Ref. \cite{Abbaslu:2024hep}.

\section{ NSI impact on \texorpdfstring{\\}{ } neutrino scattering off nuclei}\label{NSI-impact}

The interaction of neutrinos off nuclei can be parameterized as the following product of the hadronic and leptonic neutral currents
\begin{align} 
	\frac{G_{\rm F}}{\sqrt{2}}\left[\overline{\nu}_\alpha \gamma^\mu (1-\gamma_5)\nu_\beta\right](J_{\rm had}^\mu)_{\alpha \beta}.
	\label{eq:JJ}
\end{align}
The hadronic current can be decomposed as  vector and axial components
\begin{align}
	(J_{\rm had}^\mu)_{\alpha\beta}=(V_{\rm NC}^\mu)_{\alpha\beta}+(A_{\rm NC}^\mu)_{\alpha\beta}\ .
	\label{eq:Jahed}	
\end{align}
Setting $\epsilon^{Vq}=0$, $V^\mu_{NC}$ will be equal to that in the standard model. Defining $Q=(u \ d)^T$ and $\tau^3=\rm{diag} ({1,-1})$, we can write
$(V_{\rm SM}^\mu)_{\alpha\beta}$ as
\begin{align} 
	\left[(1-2\sin^2\theta_W) \overline{Q} \gamma^\mu \frac{\tau^3}{2} Q-\frac{\sin^2\theta_W}{3} \overline{Q}\gamma^\mu Q	\right] \delta_{\alpha\beta} \ .
	\label{eq:VSM}
\end{align}
However, with a nonzero $\epsilon^{Aq}$, the axial current will receive an NSI contribution:
$$ A_{\rm NC}^\mu=A_{\rm SM}^\mu +A_{\rm NSI}^\mu \  ,$$
where
\begin{align} 
	(A_{\rm SM}^\mu)_{\alpha\beta}= \left[-\overline{Q} \gamma^\mu\gamma^5 \frac{\tau^3}{2} Q\right] \delta_{\alpha\beta}
	\label{eq:ASM}
\end{align}
and $	(A^\mu_{\rm NSI})_{\alpha \beta}$ is given by
\begin{align}
\frac{\epsilon^{Au}_{\alpha \beta} +\epsilon^{Ad}_{\alpha \beta}}{2}\overline{Q}\gamma^\mu \gamma^5 Q+ \frac{\epsilon^{Au}_{\alpha \beta} -\epsilon^{Ad}_{\alpha \beta}}{2}\overline{Q}\gamma^\mu  \gamma^5\tau^3 Q . 
	\label{eq:ANSI}
\end{align}
Not surprisingly, in the presence of non-zero $\epsilon^{Aq}$, the number of neutral current scattering events at the neutrino detectors can deviate from the standard model prediction \cite{Abbaslu:2023vqk, Abbaslu:2024jzo,Abbaslu:2025fwt}. As a result, the present neutral current scattering data from MINOS($+$) as well as that from future DUNE can shed light on these parameters. The neutrino beam in both these experiments is broad band, covering all the scattering regimes of quasi-elastic, resonance and deep inelastic scattering (DIS). It is straightforward to compute the corrections to the DIS cross section induced by NSI. As shown in \cite{Abbaslu:2023vqk}, it is enough to replace the neutral couplings of the $u$ and $d$ quarks ($g^{Au}=1/2 \delta_{\alpha \beta}$ and $g^{Ad}=-1/2 \delta_{\alpha \beta}$), respectively, with 
$1/2 \delta_{\alpha \beta}+\epsilon^{Au}_{\alpha \beta}$ and $-1/2 \delta_{\alpha \beta}+\epsilon^{Ad}_{\alpha \beta}$.
However, the corrections to the resonance scattering and to the quasi-elastic scattering is more subtle. In Ref. \cite{Abbaslu:2024jzo}, we have discussed in detail how to compute the correction from the NSI to the neutral current scattering cross section.
The dominant resonance mode is the $\Delta$ resonance, $\nu+ p({\rm or} \ n) \to \nu +\Delta^+ ({\rm or} \ \Delta^0)$. Considering that the isospins of $\Delta$ and nucleons are respectively $3/2$ and $1/2$, this resonance cannot take place in the isospin singlet case, $\epsilon^{Au}=\epsilon^{Ad}$.

In general, the effect of NSI  on the cross sections of quasi-elastic and resonance scatterings appears as a correction in the relevant form factor.
For example, the cross section of the quasi-elastic scattering is given by form factors $(\tilde{F}_A^p)_{\alpha \beta}$ and  $(\tilde{F}_A^n)_{\alpha \beta}$
which in the presence of NSI are modified as
\begin{widetext}
\begin{align}
	(\tilde{F}^p_A)_{\alpha \beta} &= \left( -\frac{\delta_{\alpha \beta}}{2} + \frac{\epsilon^{A u}_{\alpha \beta} - \epsilon^{A d}_{\alpha \beta}}{2} \right) F_A + \frac{3}{2} (\epsilon^{A u}_{\alpha \beta} + \epsilon^{A d}_{\alpha \beta}) F^{(8)}_A + \left( \frac{\delta_{\alpha \beta}}{2} + \epsilon^{A u}_{\alpha \beta} + \epsilon^{A d}_{\alpha \beta}  \right) F^s_A, \label{eq:FAp}\\
	(\tilde{F}^n_A)_{\alpha \beta} &= \left( \frac{\delta_{\alpha \beta}}{2} - \frac{\epsilon^{A u}_{\alpha \beta} - \epsilon^{A d}_{\alpha \beta}}{2} \right) F_A + \frac{3}{2} (\epsilon^{A u}_{\alpha \beta} + \epsilon^{A d}_{\alpha \beta}) F^{(8)}_A + \left( \frac{\delta_{\alpha \beta}}{2} + \epsilon^{A u}_{\alpha \beta} + \epsilon^{A d}_{\alpha \beta}  \right) F^s_A.
	\label{eq:FAn}
\end{align}
\end{widetext}
For the full formulas, see Ref. \cite{Abbaslu:2024jzo}. The definition of the form factors, $F_A$, $F_A^{(8)}$ and $F_A^s$,  can be also found in \cite{Simons:2022ltq,Park:2021ypf,A1:2013fsc}. At the low energies where the quasi-elastic and resonance scatterings dominate over DIS, perturbative  methods do not apply. To compute
the cross section, we need the values of the non-perturbative  form factors $F_A$, $F_A^{(8)}$ and $F_A^s$ as inputs. There are alternative methods to extract the values of these form factors.  As we discuss in detail in \cite{Abbaslu:2024jzo}, some of these methods are valid in the presence of NSI, too. In particular,  the lattice QCD predictions for the form factors are not contaminated with NSI.
Another method to extract the form factors is based on studying the charged current interactions which again is not contaminated by the neutral current NSI.
Notice however that this method relies on the isospin symmetry to relate the form factors of charged current to those of neutral current. The third method is based on studying the neutral current interactions of neutrinos with nuclei.
Of course, in the presence of NSI, this method does not give the true values of the form factors. For our analysis, we take this issue into account.
In the presence of the lepton flavor conserving NSI, $\epsilon^{Aq}_{\alpha \alpha}$, the amplitudes of the SM contribution to the process $\nu_\alpha +{\rm nucleus} \to \nu_\alpha+X$ and that from the NSI sum up  in the amplitude level. Depending on the sign of $\epsilon^{Aq}_{\alpha \alpha}$, the interference between  the SM and NSI contributions may enhance or suppress the cross section. However, since the SM is lepton flavor conserving, there will be no contribution from the SM to $\nu_\alpha+{\rm nucleus} \to \nu_\beta +X$ with $\alpha \ne \beta$ and its contribution from $\epsilon_{\alpha \beta}^{Aq}$ will only enhance the total neutral current events. Due to the neutrino flavor oscillation, in the far detector of long baseline experiments, the beam is not a pure flavor state but a coherent linear combination of different flavors \cite{Abbaslu:2023vqk}:
\begin{widetext}
\begin{align}
	|\nu_{\rm far}(E_\nu)\rangle =\sum_i \sum_\beta e^{i m_{Mi}^2L/(2E_\nu)}(U_{\mu i}^M)^* U_{\beta i}^M |\nu_\beta\rangle \equiv \sum_\beta\mathcal{A}_\beta|\nu_\beta \rangle  \quad {\rm (neutrino\ mode)}
\end{align} 
and
\begin{align}
	|\overline{\nu}_{\rm far}(E_\nu)\rangle =\sum_i \sum_\beta e^{i \overline{m}_{Mi}^2L/(2E_\nu)}(\overline{U}_{\mu i}^M)^*\overline{U}_{\beta i}^M |\overline{\nu}_\beta\rangle \equiv \sum_\beta\overline{\mathcal{A}}_\beta|\overline{\nu}_\beta \rangle  \quad {\rm (antineutrino\ mode)} 
\end{align}
\end{widetext}
in which $U_{\beta i}^M$  and $\overline{U}_{\beta i}^M$ are the effective mixings  and $m_{Mi}$ and $\overline{m}_{Mi}$ are respectively the  mass eigenvalues of neutrinos and antineutrinos in matter.  $\mathcal{A}_\beta$ and 
$\overline{\mathcal{A}}_\beta$ are the oscillation amplitudes of neutrino and antineutrino:
$$|\mathcal{A}_\beta|^2=P(\nu_\mu\to \nu_\beta)  \ \ {\rm and} \ \ |\overline{\mathcal{A}}_\beta|^2=P(\overline{\nu}_\mu\to \overline{\nu}_\beta)\ .$$
The amplitudes of the inclusive  scatterings of $\nu_{far}$ and  $\bar{\nu}_{far}$ off 
nuclei to a given (anti)neutrino flavor, $\alpha$, can be written as
\begin{widetext}
\begin{align} \label{MM}
	\mathcal{M}(\nu_{\rm far} +{\rm nucleus}\to \nu_\alpha +X)=& \sum_\beta \mathcal{A}_\beta \mathcal{M}(\nu_{\beta} +{\rm nucleus} \to \nu_\alpha +X), \notag \\
	\mathcal{M}(\overline{\nu}_{\rm far} +{\rm nucleus} \to \overline{\nu}_\alpha +X)=& \sum_\beta \overline{\mathcal{A}}_\beta  \mathcal{M}(\overline{\nu}_{\beta} +{\rm nucleus} \to \overline{\nu}_\alpha +X),
\end{align}
\end{widetext}
in which $X$ can be any state.
Notice that because of the oscillation, even for the case of off-diagonal lepton flavor violating NSI, $\epsilon^{Aq}_{\alpha \beta}$ $\alpha \ne \beta$,  constructive or destructive interference with the standard model contribution can take place. As shown in Refs. \cite{Abbaslu:2023vqk,Abbaslu:2025fwt},
with proper changes of flavor basis, the formulas can be very simplified.
Unlike the vector NSI, the axial NSI does not affect the neutrino propagation in matter. Thus, $\mathcal{A}_\beta$ are given by the same formulas as in the standard model.

\section{\label{exps} MINOS, MINOS+ and DUNE }
The DUNE experiment is a broad band long baseline experiment which is planned to start its neutrino beam in 2031 \cite{DUNE:2020ypp,DUNE:2021cuw}. The neutrino beam will be located in Fermilab.  DUNE will be equipped with two detectors: 1) Near Detector (ND) and 2) Far detector  (FD). These two detectors are respectively located at 574 m and at 1297 km downstream from the beginning of horn 1 \cite{DUNE:2021cuw}.
The detectors of DUNE will have a state-of-the-art technology with remarkable energy and spatial resolution.

These detectors can distinguish the charged current from the neutral current events.
The full capabilities of the DUNE detectors are yet to be explored. With developing neural network techniques, the capabilities may extend to areas that had not been envisaged in the original design. One possibility is to distinguish between DIS and resonance scattering with the methods discussed in \cite{Tingey:2022evd}. In Ref. \cite{Abbaslu:2023vqk}, we have shown how by studying the neutral current deep inelastic scattering events in DUNE, information on the axial NSI can be derived.
In this analysis, the signal is made of  the neutral current scattering  events and the background is mainly composed of misidentified charged current or resonant neutral current scattering events with misidentification probability of $10 \%$
\cite{Coloma:2017ptb}.
We take fiducial masses for ND and FD, respectively, equal to 67.2 ton and 40 kton \cite{DUNE:2020ypp,DUNE:2021cuw}. 
The detectors are made of Argon. The unoscillated flux has been taken from  Ref.~\cite{site}.
We have assumed $1.1 \times 10^{21}$ protons on target (POT) per year and 6.5 years of data taking in each neutrino and antineutrino modes.
In the neutral current scattering events, the final neutrino, appearing as missing energy and momentum in the detector, carries away a significant and unknown amount of the energy and momentum of the initial neutrino.
To our
  best knowledge, the migration matrices for the neutral current events at DUNE are yet to be developed. We have therefore only considered the total number of neutral current scattering events, without binning the mock data.
We have included two sources of uncertainty: the normalization uncertainty ($\sigma_\epsilon$) and the uncertainty in the determination of the background, ($\sigma_\omega$). We have treated them with the pull-parameter method. For more details about the analysis, see Ref. \cite{Abbaslu:2023vqk}.

Let us now describe the MINOS and MINOS+ experiments. Both these experiments used the  NUMI beam from Fermilab and two detectors made of steel planes and scintillator strips. The distance to the near detector was 1.04 km and that to the far detector was 735 km \cite{MINOS:2017cae}. The 
MINOS experiment has run from 2005 to 2012 with a peak energy at 3~GeV. MINOS$+$ started at 2013 with a peak energy at 7~GeV.
Along with the charged current events which were used to measure the parameters of the standard three neutrino oscillation scheme,  MINOS($+$) collected neutral current scattering data. In Ref. \cite{MINOS:2017cae}, the MINOS collaboration has used this data to constrain the $3+1$ neutrino scheme. In Ref. \cite{Abbaslu:2025fwt}, we used the information in the ancillary files of \cite{MINOS:2017cae} to constrain the axial NSI.
 Both the near and far detector neutral current data include 27 energy bins spanning from 0 to 40 GeV. Using the covariance, $V$, given in \cite{MINOS:2017cae} , we can define the $\chi^2$ statistic as
\begin{equation}  \label{ChiNC}
	\chi^2_{NC}=\sum_{i=1}^{27} \sum_{j=1}^{27} (\mathcal{N}_i^{obs}-\mathcal{N}_i^{pre}(\epsilon))(V^{-1})_{ij}(\mathcal{N}_j^{obs}-\mathcal{N}_j^{pre}(\epsilon)) \ .
\end{equation}
$\mathcal{N}_i^{obs}$ denotes the observed NC events in the $i$th bin and $\mathcal{N}_i^{pre}(\epsilon)$ is the prediction for the  $i$th bin as a function of the  NSI coupling $\epsilon$. $\mathcal{N}_i^{pre}(\epsilon)$
 includes both the signal and the background. In \cite{Abbaslu:2025fwt}, we explain how to compute $\mathcal{N}_i^{pre}(\epsilon)$  in detail.

To compute the cross sections, we have used the NuWro Monte Carlo neutrino event generator \cite{nuwro2025, Juszczak:2005zs, Golan:2012rfa, Golan:2012wx}. We have implemented the necessary changes to include the axial NC NSI effects.  NuWro takes into account the nuclear effects which are particularly important for the quasi-elastic and resonance regimes. While for the electromagnetic form factors of the quasi-elastic scattering, we have used the BBBA05 parameterization method \cite{Bradford:2006yz},  for the axial form factors, we used the dipole forms with the parameters listed in Table 1 of Ref \cite{Abbaslu:2024jzo}. For the resonance scattering, we have taken the dipole delta form factors with parameters given in Table 2 of Ref \cite{Abbaslu:2024jzo} as inputs  for the NuWro event generator.
 
\section{\label{results} Results }
\begin{figure*}[htb]
	\centering
	\subfigure[]{\includegraphics[width=0.48\textwidth ]{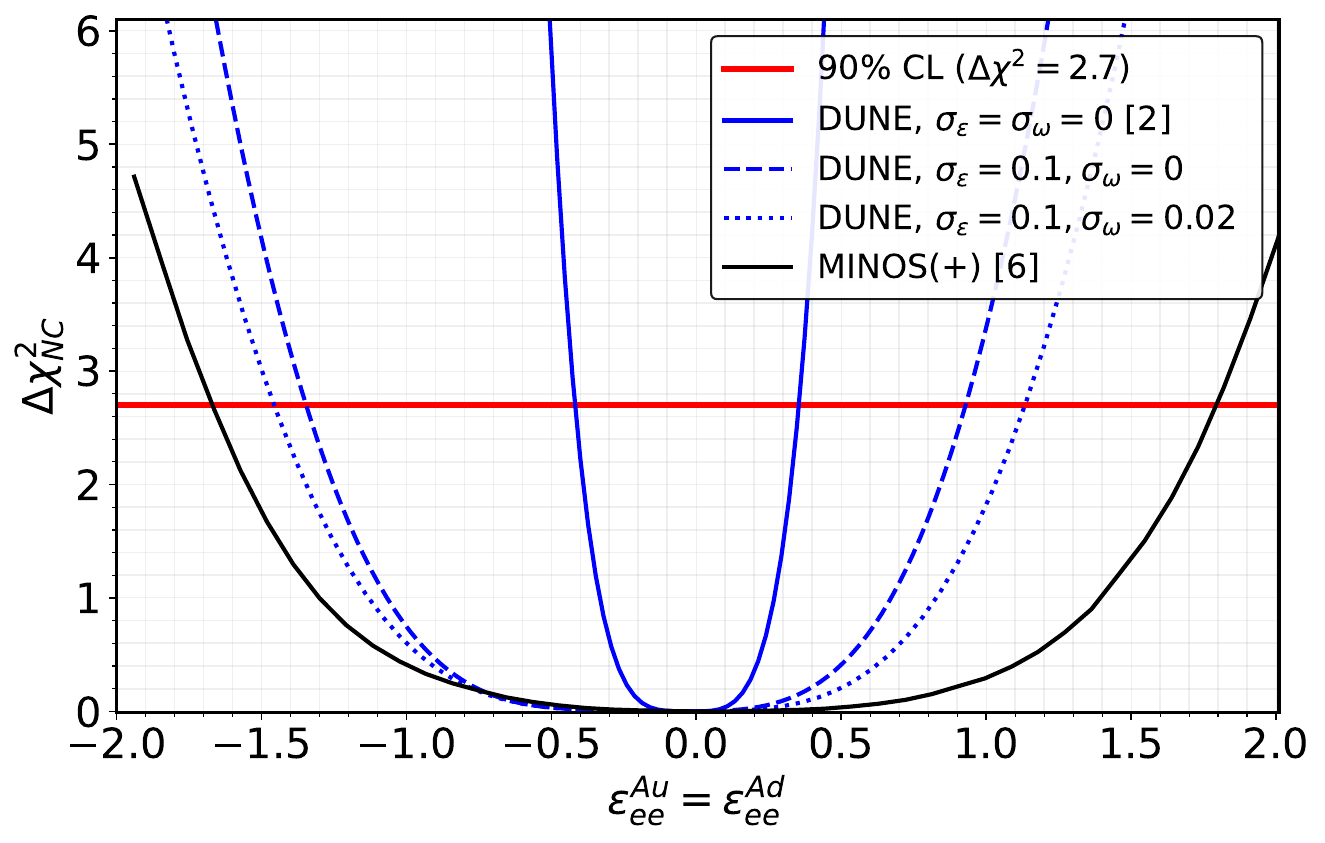}\label{fig:delta-chi2-ee-Au=Ad}}
	\subfigure[]{\includegraphics[width=0.48\textwidth ]{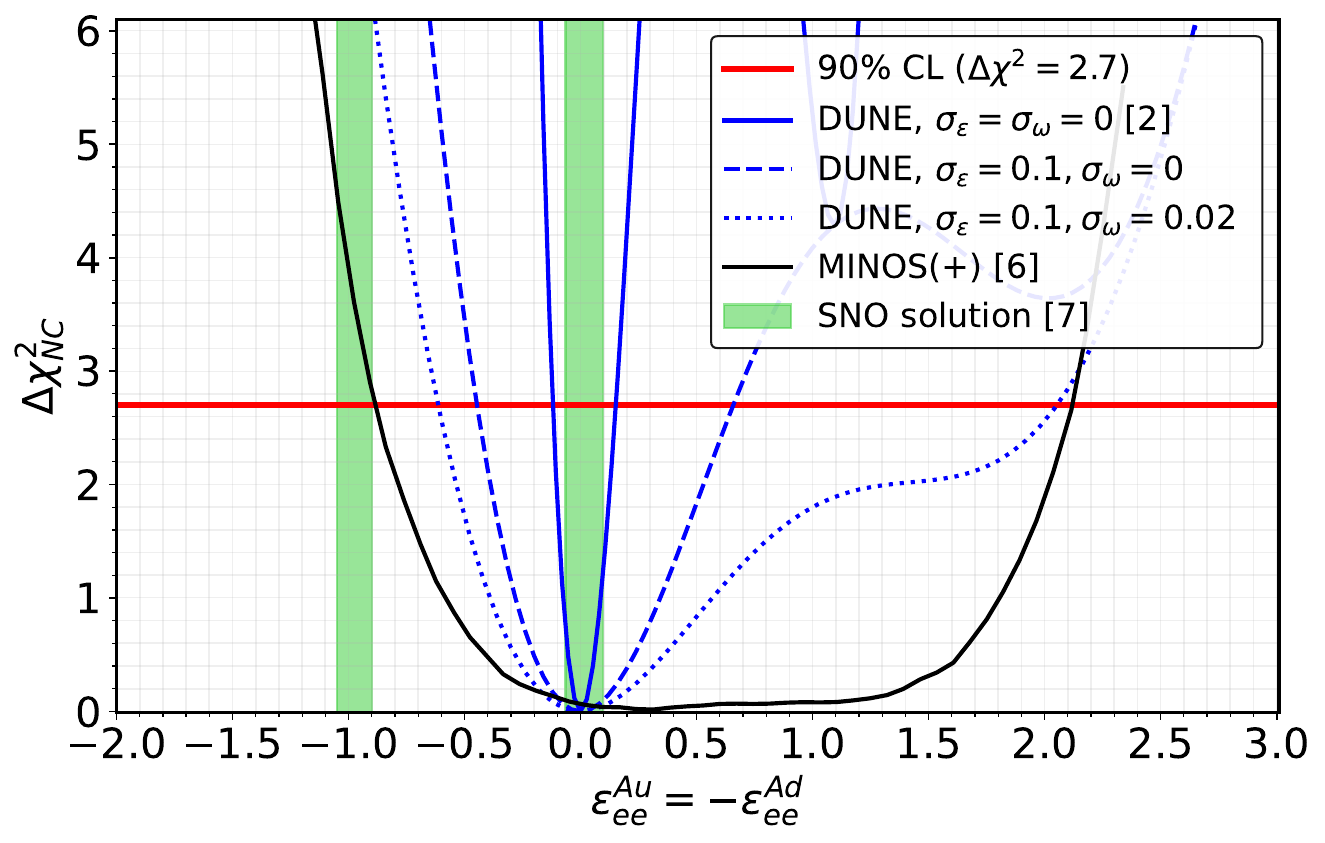}\label{fig:delta-chi2-ee-Au=-Ad}}
	\subfigure[]{\includegraphics[width=0.48\textwidth ]{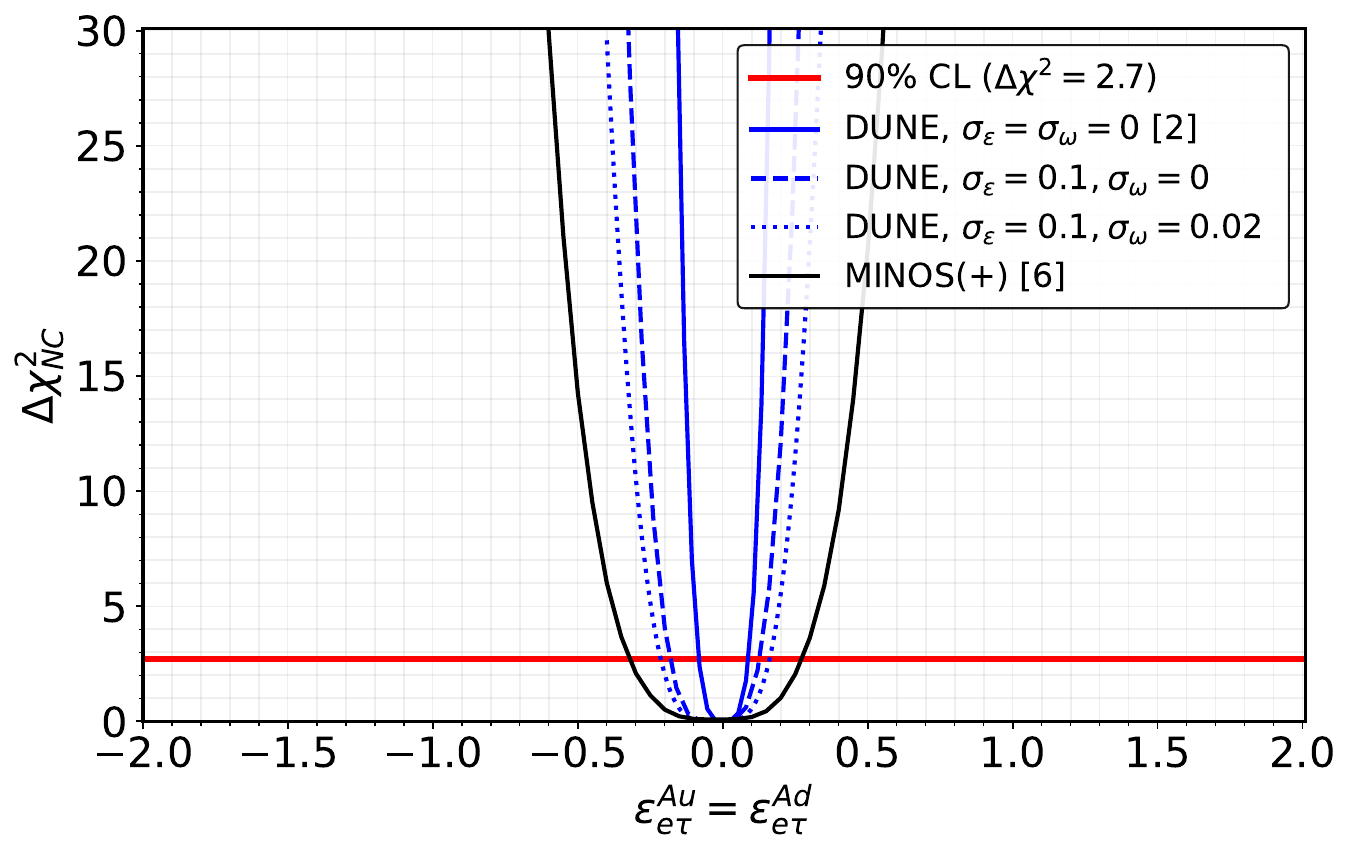}\label{fig:delta-chi2-et-Au=Ad}}
	\subfigure[]{\includegraphics[width=0.48\textwidth ]{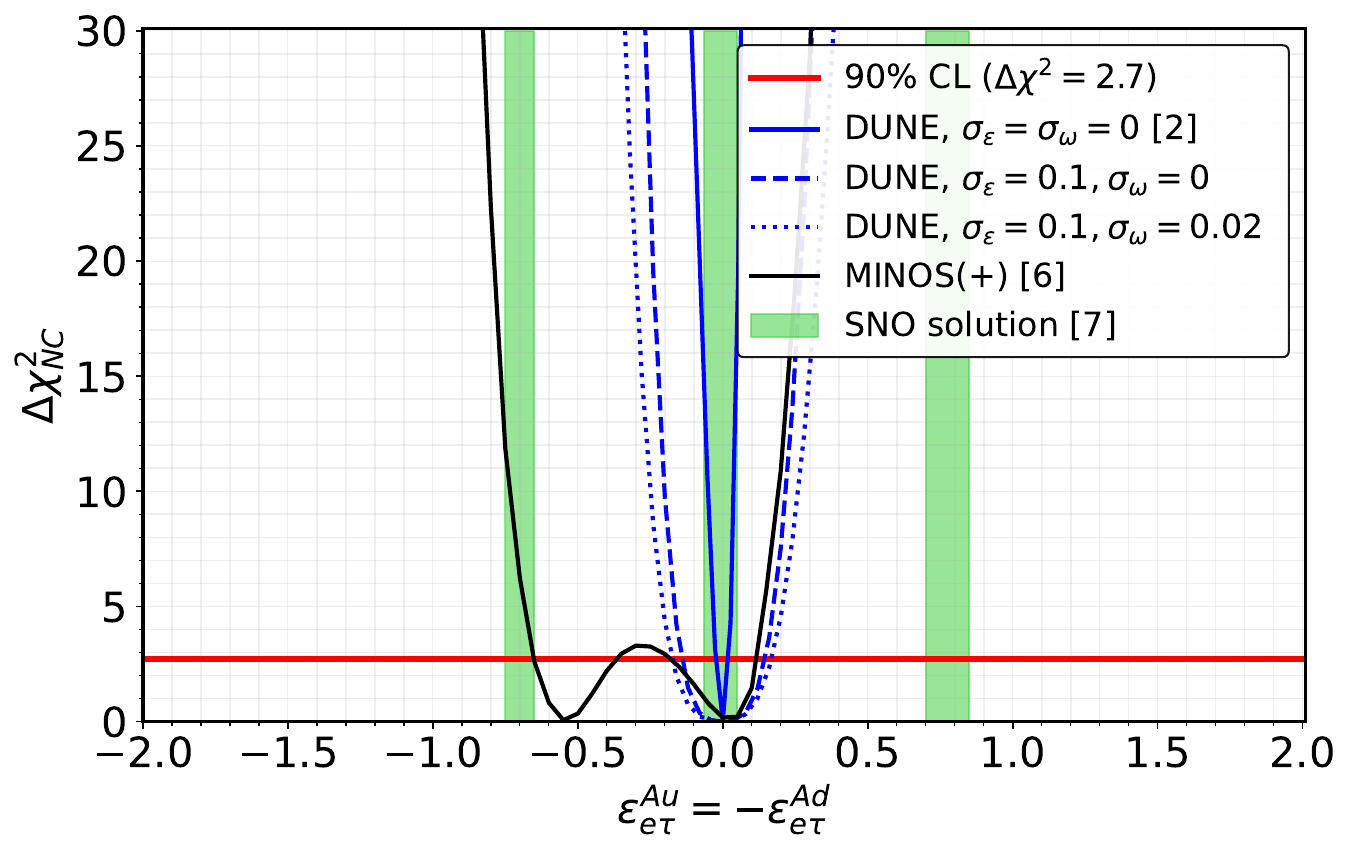}\label{fig:delta-chi2-et-Au=-Ad}}
	\subfigure[]{\includegraphics[width=0.48\textwidth ]{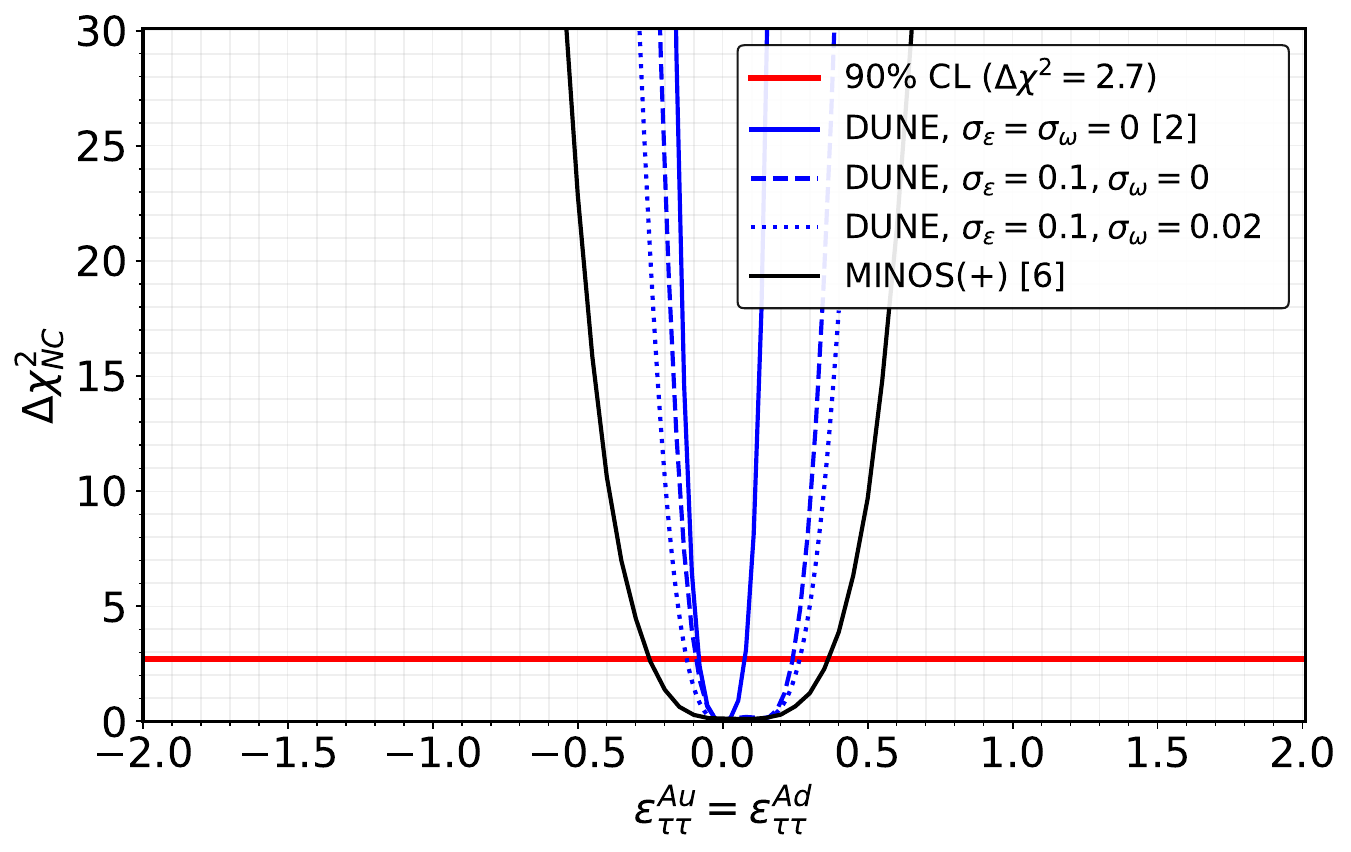}\label{fig:delta-chi2-tt-Au=Ad}}
	\subfigure[]{\includegraphics[width=0.48\textwidth ]{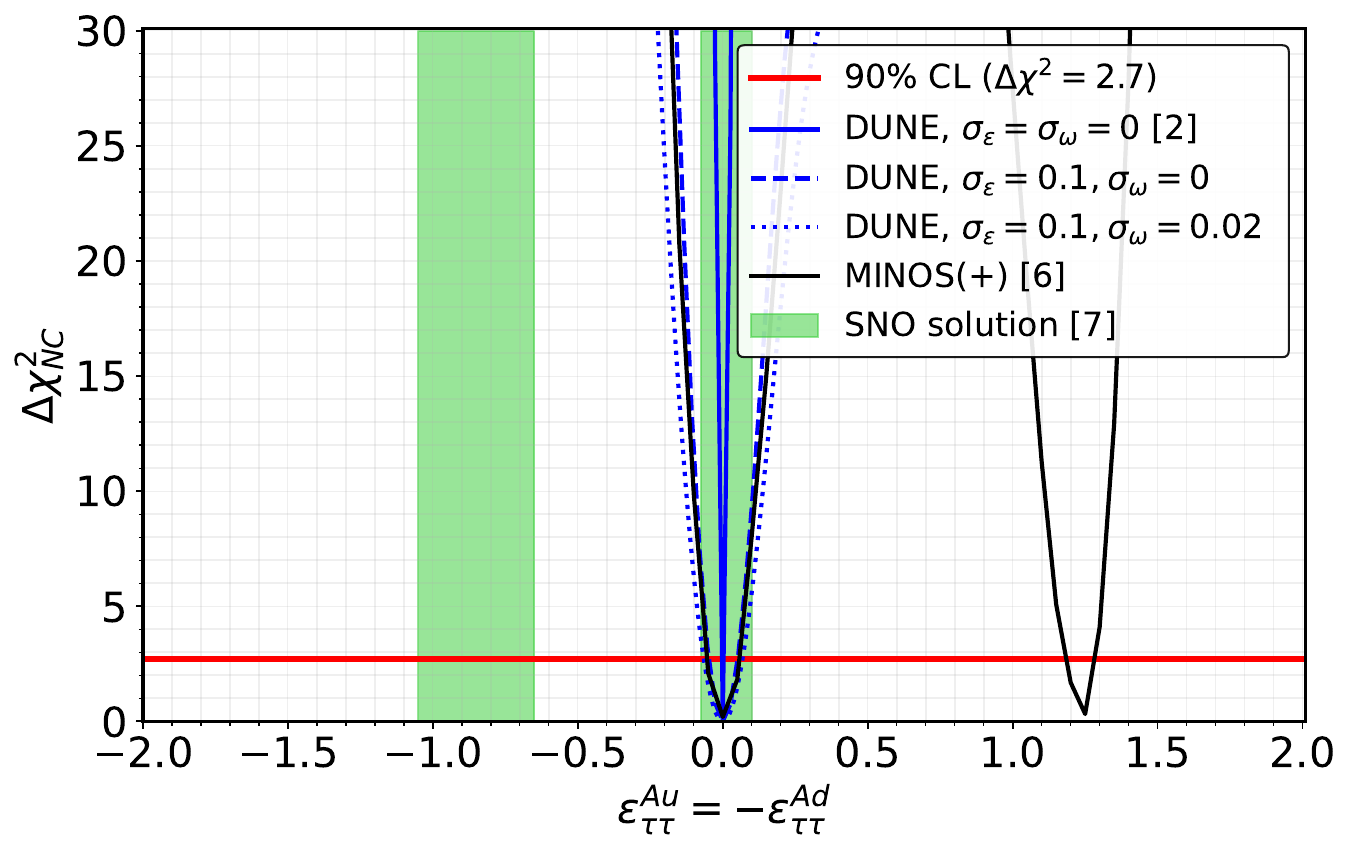}\label{fig:delta-chi2-tt-Au=-Ad}}
	\caption{ $\Delta\chi^2_{NC}$ versus the NSI couplings for the isospin
		singlet (left) and the isospin violating couplings (right) for the $ee$, $e\tau$, and $\tau\tau$ components. In each panel, only one NSI coupling is set nonzero. The horizontal red lines show $\Delta \chi^2_{NC}=2.7$, which corresponds to 90 \% C.L. with one degree of freedom. The blue solid lines depict the bounds forecasted for the DUNE experiment assuming zero uncertainty \cite{Abbaslu:2023vqk}. The dashed (dotted) blue curves show the forecasted bounds for the DUNE experiment with $\sigma_{\epsilon}=0.1$ and $\sigma_{\omega}=0$ (with $\sigma_{\epsilon}=0.1$ and $\sigma_{\omega}=0.02$) \cite{Abbaslu:2023vqk}. The solid black lines show the bounds based on the MINOS and MINOS+ NC data \cite{Abbaslu:2025fwt}. The green bands represent the 90 \% C.L. bounds obtained by dominated the SNO neutral current data \cite{Coloma:2023ixt}. }
	\label{Fig:delta-chi2}
\end{figure*}

\begin{table*}[t]
	\renewcommand{\arraystretch}{0.7}
	\begin{tabular*}{\textwidth}{@{\extracolsep{\fill}}lcccl@{}}
		Coupling&Source& 90\% C.L bound \\
		\hline
		\multirow{3}{*}{$\epsilon_{ee}^{Au}$=$\epsilon_{ee}^{Ad}$} & Based on SNO NC data~\ \cite{Coloma:2023ixt}& - \\ 
		& DUNE  \cite{Abbaslu:2023vqk}& [-0.41 , 0.35 ]\\ 
		&MINOS(+) \cite{Abbaslu:2025fwt} & [-1.67 , 1.8] \\
		\hline
		\multirow{3}{*}{$\epsilon_{e\tau}^{Au}$=$\epsilon_{e\tau}^{Ad}$} & Based on SNO NC data~\ \cite{Coloma:2023ixt}& -  \\ 
		& DUNE  \cite{Abbaslu:2023vqk}& [-0.083 , 0.089 ] \\ 
		& MINOS(+)  \cite{Abbaslu:2025fwt} & [-0.32 , 0.27] \\
		\hline
		\multirow{3}{*}{$\epsilon_{\tau\tau}^{Au}$=$\epsilon_{\tau\tau}^{Ad}$} & Based on SNO NC data~\ \cite{Coloma:2023ixt}&-  \\ 
		& DUNE  \cite{Abbaslu:2023vqk}& [-0.084 , 0.076 ] \\ 
		& MINOS(+) \cite{Abbaslu:2025fwt} & [-0.25 , 0.36] \\ 
		\hline\hline
		\multirow{3}{*}{$\epsilon_{ee}^{Au}$=-$\epsilon_{ee}^{Ad}$} & Based on SNO NC data~\ \cite{Coloma:2023ixt}& $ [-1.05,-0.9]+[-0.095, 0.065]$ \\ 
		& DUNE  \cite{Abbaslu:2023vqk}& [-0.12 , 0.15]\\ 
		&MINOS(+)  \cite{Abbaslu:2025fwt} & [-0.88 , 2.12] \\
		\hline
		\multirow{3}{*}{$\epsilon_{e\tau}^{Au}$=-$\epsilon_{e\tau}^{Ad}$} & Based on SNO NC data~\ \cite{Coloma:2023ixt}&$[-0.75, -0.65]+[-0.065, 0.05]+[+0.7, 0.85]$  \\ 
		& DUNE  \cite{Abbaslu:2023vqk}& [-0.023 , 0.017] \\ 
		&MINOS(+) \cite{Abbaslu:2025fwt} & [-0.65 , -0.37]$+$[-0.17 , 0.12] \\
		\hline
		\multirow{3}{*}{$\epsilon_{\tau\tau}^{Au}$=-$\epsilon_{\tau\tau}^{Ad}$} & Based on SNO NC data~\ \cite{Coloma:2023ixt}& $[-1.05,-0.9]+[-0.095,+0.075] $  \\ 
		& DUNE  \cite{Abbaslu:2023vqk}& [-0.003 , 0.003] \\ 
		&MINOS(+) \cite{Abbaslu:2025fwt} & [-0.05 , 0.05]$+$[1.18 , 1.28]\\ 
		\hline
	\end{tabular*}
	\caption{The 90\% C.L. bounds on the $ee$, $e\tau$, and $\tau\tau$ components  for the  isospin singlet ($\epsilon^{Au}=\epsilon^{Ad}$) and isospin violating ($\epsilon^{Au}=-\epsilon^{Ad}$) NSI configurations.} \label{tab:bounds}
\end{table*}

In Fig \ref{Fig:delta-chi2}, you can see the bounds on the different flavor components of the axial NSI for both the isospin singlet $(\epsilon^{Au}=\epsilon^{Ad}$) and for the isospin violating configurations  $(\epsilon^{Au}=-\epsilon^{Ad}$). 
The black curve shows our results found in \cite{Abbaslu:2025fwt} using the neutral current data collected by MINOS($+$).
Through $\mathcal{A}_\beta$ in Eq. (\ref{MM}), the rate of the neutral current interactions in the presence of NSI at the far detector will depend on the neutrino mass and mixing parameters. For MINOS($+$), the dependence of the rate on $\theta_{12}$ and $\Delta m_{21}^2$ is mild because $\Delta m_{21}^2 L/(4E_\nu)\ll 1$. We have therefore fixed them to the central best fit value in \cite{Abbaslu:2025fwt}; however, we have marginalized over the rest of neutrino mass and mixing parameters.

{We have employed a Markov Chain Monte Carlo (MCMC) analysis using the Cobaya framework with a Metropolis-Hastings sampler to probe the parameter space of standard neutrino oscillation parameters and non-standard interaction (NSI) parameters.   In the MCMC analysis, we set  $\theta_{12}$ and $\Delta m_{21}^2$ to their global best-fit values while applying the Gaussian priors to the remaining standard neutrino oscillation parameters, with the central values and widths (1$\sigma$ uncertainties) given in Table~1 of Ref \cite{Abbaslu:2025fwt}.
	More details are presented in Ref \cite{Abbaslu:2025fwt}.}

To obtain bounds on NSI couplings of $\epsilon^{Aq}$, we use $\Delta \chi^2$ defined by
\begin{equation*}
  \Delta\chi^{2}_{NC} (\epsilon)\equiv \chi^{2}_{NC} (\epsilon)- \chi^{2}_{NC}|_{min},
\end{equation*}
where $\chi^{2}_{NC}|_{min}$ is the  minimum varying over nonzero $\epsilon$ and neutrino mass and mixing parameters. The vertical axes of Fig.~\ref{Fig:delta-chi2}
show $  \Delta\chi^{2}_{NC} $ as a function of the nonzero NSI component indicated by the label of the horizontal axes, setting the rest of the NSI couplings equal to zero and marginalizing over the neutrino mass and mixing parameters. In Ref.~\cite{Abbaslu:2025fwt}, we have also studied the case that more than one lepton flavor component of $\epsilon^{Aq}$ is nonzero.

 We have also superimposed the forecasted bound for the DUNE experiment which we have found in Ref. \cite{Abbaslu:2023vqk}.
The solid blue curve shows the bound in the ideal case of zero systematic uncertainties. The dashed lines show the bounds from the  DUNE experiment taking a normalization uncertainty of $\sigma_\epsilon =10 \%$  and zero uncertainty in the background (mis)identification $\sigma_\omega=0$. The  	 dotted lines show the same with both normalization and background (mis)identification uncertainties, taking  $\sigma_\epsilon =10 \%$ and $
\sigma_\omega =2 \%$.
	 In the $\epsilon^{Au}=-\epsilon^{Ad}$ panels, the vertical green regions show the solutions in \cite{Coloma:2023ixt} which is dominated by SNO. As seen from these figures, MINOS($+$) already rules out the disconnected nonzero solution.  On the other hand, SNO rules out the degenerate solution to the MINOS($+$) neutral current data. Thus, the MINOS and SNO results are complementary.

As seen from Fig.~\ref{Fig:delta-chi2} , the MINOS($+$) bound on $ee$ is still very weak allowing $\epsilon^{Aq}_{ee}\sim 1$. Even the DUNE experiment cannot significantly improve this bound. The bound on $\epsilon^{Aq}_{ee}$ is statistically limited because $P(\nu_\mu \to \nu_e)$ is suppressed by $\sin^2 \theta_{13}$.
However, the MINOS($+$) data have significantly improved the bound on the $\tau \tau$ and $e \tau$ components. This is understandable because at the far detectors of both MINOS($+$) and DUNE, $P(\nu_\mu \to \nu_\tau)\sim 1$. 

Fig.~\ref{Fig:delta-chi2} demonstrates that even with a conservative assumption on the systematic uncertainties, DUNE can improve the present bound from MINOS($+$) on $\epsilon^{Au}_{e\tau}=\epsilon^{Ad}_{e\tau}$ and $\epsilon^{Au}_{\tau\tau}=\epsilon^{Ad}_{\tau\tau}$.  In the case of $\epsilon^{Au}=-\epsilon^{Ad}$, the possibility of improvement by DUNE will be limited by the systematic uncertainties. Table \ref{tab:bounds} shows the bounds on different components.

\section{Summary \label{Con}}
We have first briefly described the model proposed in \cite{Abbaslu:2024hep} which leads to $\epsilon^{Au}_{\tau \tau}=\epsilon^{Ad}_{\tau \tau}$, demonstrating that large axial NSI may be a window to rich new physics. We have then discussed whether the existing neutral current scattering data from MINOS and MINOS$+$ can improve the bounds on the axial NSI. We have presented the bounds on $\epsilon_{\tau\tau}^{Aq}$,  $\epsilon_{e\tau}^{Aq}$ and $\epsilon_{ee}^{Aq}$ both in the isospin singlet and isovector cases. Since at the far detector of DUNE, the beam is mostly composed of $\nu_\tau$, the bound on $\epsilon^{Aq}_{\tau \tau}$ can be significantly improved; see Table \ref{tab:bounds}. However,  since the $\nu_e$ component of the flux at the far detector is suppressed by $\sin^2 \theta_{13}$, the statistics is not enough to constrain the $ee$ components.

Before \cite{Abbaslu:2025fwt}, the analysis of the available data (including the SNO neutral current dissociation data but without the MINOS neutral current data) had found non-trivial disconnected solutions with nonzero $\epsilon^{Au}_{\tau \tau}=-\epsilon^{Ad}_{\tau \tau}$,  $\epsilon^{Au}_{e \tau}=-\epsilon^{Ad}_{e \tau}$ and  $\epsilon^{Au}_{ee}=-\epsilon^{Ad}_{ee}$. The MINOS($+$)
neutral current completely rules out these solutions; see Fig.~\ref{Fig:delta-chi2}. We have also shown that in the future,  DUNE can significantly improve the bounds even in the most pessimistic case of large normalization and background misidentification uncertainties.

\begin{acknowledgments}
This article is prepared for the proceedings of  the 22nd Lomonosov conference which was held during 21-27 August 2025. Y.F. would like to thank the organizers, especially Prof. Alexander I. Studenikin, for the kind invitation.
\end{acknowledgments}











\section*{FUNDING}
This project has received funding from the European Union’s Horizon Europe research and innovation programme under the Marie Skłodowska-Curie Staff Exchange grant agreement No 101086085 – ASYMMETRY. The authors acknowledge the computational resources provided by the SARV computing facility at the school of theoretical physics of IPM.

\section*{CONFLICT OF INTEREST}
The authors declare that they have no conflicts of interest.



\end{document}